\documentstyle[art10,hip-100ps,twoside,epsf]{article}
\def\D{\Delta}

\def\Dt{\Delta\tau}

\def\t0{\tau_0}

\def\t{{\tau}}

\def\ben{\begin{eqnarray}}
\def\enn{\end{eqnarray}}
\def\ov{\over\displaystyle\strut}
\def\dst{\displaystyle\phantom{\mid}}
\def\l({\left(}
\def\r){\right)}
\def\o{{out}}
\def\s{{side}}

\begin{document}
\cimo{221}{232} \setcounter{page}{221} 
\thispagestyle{empty} \vskip 1cm 
\noindent
{\Large\bf
	Hydrodynamic Parameterization of NA44 Data
	on Kaon and Pion Correlations and Spectra
}\\[3mm] 
\def\rightmark{Hydrodynamic Parameterization of NA44 Data 
  }\def\leftmark{T. Cs{\"o}rg\H o and B. L{\"o}rstad}
\hspace*{6.327mm}\begin{minipage}[t]{12.0213cm}{\large\lineskip .75em
T. Cs{\"o}rg\H o$^{1,2a}$ and B. L{\"o}rstad$^{3b}$
}\\[2.812mm] 
\hspace*{-8pt}$^1$ Department of Physics, Columbia University \\
538 W 120th St New York, N. Y. 10027 \\
\hspace*{-8pt}$^2$ KFKI Research Institute for Particle and Nuclear Physics \\
H-1525 Budapest 114, POB 49, Hungary  \\
\hspace*{-8pt}$^3$ Physics Department, Lund University\\
S-221 00 Lund, POB 118, Sweden
\\[4.218mm]{\it
Received 30 June 1996 
}\\[5.624mm]\noindent
{\bf Abstract.}
An analytically calculable model~\cite{tamas2} has been proposed recently
for the simultaneous description of particle spectra and correlations,
which is able to explain in a natural manner the $m_t$ scaling of the
radius parameters of the Bose--Einstein correlation functions.
We present here {\it preliminary} results on fitting this model  
to data sampled by NA44 collaboration in central S + Pb 
reactions at CERN SPS, which indicate that both pions and kaons 
freeze-out at a surprisingly low temperature of $T \simeq 108 $ MeV.
\end{minipage}

\section{Introduction}
 
Recently there has been much interest in the measurement and 
the calculation of the Bose--Einstein correlation functions (BECFs) 
and those of the invariant momentum distributions (IMDs) 
for rapidly expanding systems with flow and temperature profiles.
The expansion may result in strong correlations between 
space-time and momentum space variables implying that the
 Bose--Einstein correlations are in general 
not measuring the geometrical sizes of big 
and expanding finite systems.
The geometrical sizes play an important role in the description of the
invariant momentum distributions.
 That is  why a simultaneous fit of IMDs and BECFs
gives good constraints on the parameters of the model.

We will discuss an attempt to fit simultaneously the data of S+Pb interactions
analysed by the NA44 collaboration \cite{na44mt}. They showed an unexpected, scaling
behaviour for the radii parameters of the BECF, $R_{side},R_{out}$ and $R_{long}$.
These parameters turned out to be proportional to $1/\sqrt{m_t}$, where $m_t$ is the mean
transverse mass of the identical particle pair. The same proportionality factor
was valid for both pions and kaons. The NA44 collaboration has also measured
the IMDs for the same S+Pb interactions using similar trigger conditions.
Data are not yet fully normalized but that is not so serious for our purpose as
we will only fit the shape of the IMD anyway. We have two reasons to only take the shape into
account:

({\it i}) The slightly different trigger conditions 
for data sampled to measure
IMD and for BECF might give slightly different
normalizations but the shape is observed to be uneffected \cite{michael}.

({\it ii}) The expansion model 
is applicable only to the central finite core creating particles, not
to all particles measured. This core is characterized by 
hydrodynamic, collective behaviour and small lengths of homogeneity.
This is in contrast to the halo, created by the decay products of
long-lived resonances, which is characterized by a free streaming and 
decay of these resonances and correspondingly with large lengths of 
homogeneity proportional to the decay time of the long-lived resonances,
~\cite{halo}. 
The $\lambda$ parameter which measures the strength of
the correlation gives a handle to estimate the fraction of particles
which are produced in the core~\cite{schlei,halo}. 
When $\lambda$ is independent of
$m_t$ the IMD of the core particles is proportional to the IMD of all
particles measured. The S+Pb data show an approximately $m_t$ independent
$\lambda$ parameter, $\lambda_{\pi^+} = 0.56 \pm 0.02$ and $0.55 \pm 0.02$
 at the quite different mean transverse momenta of $150 $ MeV 
and $ 450 $ MeV, respectively.

We will shortly present the model of the core, present the results
of a fit to three sets of IMDs and the corresponding radii measurements using
({\it i}) an analytical approximation 
and ({\it ii}) a numerical integration method. We will summarize
with some conclusions.

\section{A Model of the Core }

For central heavy ion collisions at high energies the beam or $z$ axis
becomes a symmetry axis. Since the initial state of the reaction is
axially symmetric and the equations of motion do not break this pattern,
the final state must be axially symmetric, too.
However, in order to generate the thermal length scales in  the transverse
directions, the flow field must be either three-dimensional,
or the temperature must have significant gradients in the 
transverse directions. Furthermore, the local temperature
may change during the the duration of the particle
emission either because of the re-heating of the system caused by the
hadronization and/or intensive rescattering processes   
or the local temperature may decrease because of the expansion
and the emission of the most energetic particles from the 
interaction region.
We consider the following emission function for 
high energy heavy ion reactions as a model of the core:
\begin{eqnarray}
S_c(x,p) \, d^4 x & = & {\dst  g \ov (2 \pi)^3} \, 
	{\dst d^4 \Sigma^{\mu}(x) p_{\mu} \ov
	\exp\l({\dst  u^{\mu}(x)p_{\mu} \ov  T(x)} - 
        {\dst \mu(x) \ov  T(x)}\r) - 1} \ , \label{e:s}
\enn
where $ d^4 \Sigma^{\mu}(x) p_{\mu} $ describes the flux of particles through a 
finite layer of freeze-out hypersurfaces. We assume that this layer is
parameterized by $ \tau = \sqrt{t^2 - z^2} $ where the random variable $\tau$
is characterized by a probability distribution, such that  
\ben
	d^4 \Sigma^{\mu}(x) p_{\mu} & = & 
	 m_t \cosh[\eta - y]  \, H(\tau) d\tau \, \tau_0 d\eta \, dr_x \, dr_y \ ,
\enn
where the  finite duration is modelled by 
  $H(\tau) \propto \exp(-(\tau-\tau_0)^2 
/(2 \Delta\tau^2))$.  Here 
$\tau_0$ is the mean emission time,
        $\Delta \tau$ is the duration of the emission in (proper) time.
        The four-velocity  and the local temperature and density profile
	 of the expanding matter is given by
\ben
u(x) & = & \l( \cosh[\eta] \cosh[\eta_t],
        \, \sinh[\eta_t] {\dst r_x \ov r_t}, 
	\, \sinh[\eta_t] {\dst r_y \ov r_t}, 
	\, \sinh[\eta] \cosh[\eta_t] \r), \\
{\dst 1 \ov T(x)} & =  &
	{\dst 1 \ov T_0 } \,\,
	\left( 1 + a^2 \, {\dst  r_t^2 \ov 2 \tau_0^2} \right) \,
	\left( 1 + d^2 \, {\dst (\tau - \tau_0)^2 \ov 2 \tau_0^2  } 
	\right) , \qquad 
\sinh[\eta_t] \,\,  = \,\,  b {\dst r_t\ov \tau_0 },
	\\
	{\dst \mu(x) \ov T(x) }  & = &  {\dst \mu_0 \ov T_0} -
        { \dst r_x^2 + r_y^2 \ov 2 R_G^2}
        -{ \dst (\eta - y_0)^2 \ov 2 \Delta \eta^2 }, \label{e:mu}
\end{eqnarray}
	where $\mu(x)$ is the chemical potential and $T(x)$ is the local
        temperature characterizing the particle emission.

	Note that in Ref.~\cite{tamas2} we have considered the Boltzmann
	approximation to the emission function given by the present 
	eqs.~(\ref{e:s} - \ref{e:mu}). Thus the results obtained in
	Ref.~\cite{tamas2} correspond to Boltzmann approximations.
	Here we indicate some new analytic results for the single particle
	inclusive spectra which are obtained {\it without} the full
	Boltzmann approximation. 

	The emission function of eqs.~(\ref{e:s} - \ref{e:mu}) corresponds to 
	a {\it longitudinally expanding,  finite} system,
	with {\it transverse and temporal temperature profile} and 
	with a linear {\it transverse flow} velocity profile. 
	The decrease of the temperature in the transverse direction is
	controlled by the parameter $a$, while 
	the strength of the transverse flow is controlled by the parameter $b$.
	The parameter $d$ controls the strength of the change of the 
	local temperature during the course of particle emission.
	This parameterization corresponds to a Gaussian approximation  
	of the local flow, inverse temperature and chemical potential
	profiles, as discussed in detail in Ref.~\cite{tamas2}. For the case 
	of $a = d = 0$ we recover a model also discussed in Ref.~\cite{uli},
	while for the case of $d = 0$ the model of Ref.~\cite{qm95} 
	is reproduced, while for $ b = d = 0$ we obtain a version of a model
	proposed in Ref.~\cite{akk}.  

	Using a saddle point approximation similar to the one presented in  
	Ref.~\cite{tamas2}, but keeping the Bose-Einstein form
	of the emission function, we find that the  IMD of 
	  the particles produced from the core plus the halo 
	can be expressed as:
\ben
	{\dst d^2 n\ov dy \, dm_t^2 } & = & 
			{\dst 1 \ov \sqrt{\lambda_*} }
			{\dst d^2 n_c \ov dy \, dm_t^2 }, \\
	{\dst 1 \ov \pi} {\dst d^2 n_c \ov dy \, dm_t^2 } & = & 
	{\dst g \ov (2 \pi)^3 } \,\, 
	V_*({x}_s) \,\,
	{\dst 	p^{\mu} n_{\mu}({x}_s) \ov
		{\exp\l({\dst  u^{\mu}( {x}_s)p_{\mu} \ov 
			 T( {x}_s)} 
		- {\dst \mu( {x}_s) \ov { T( {x}_s)} }\r) - 1}} \ .
\enn
This result can be considered new as
compared to the ones presented in Ref.~\cite{tamas2},
but applying a Boltzmann approximation to this result reproduces the
expressions given in Ref~\cite{tamas2}.  Parts of this expression 
 are explained piece by piece below.  The intercept parameter $\lambda_*$, 
which relates the total number of particles
to the number of particles originating from the core,
 is the effective intercept parameter of the 
Bose--Einstein correlation function. The intercept parameter $\lambda_*$ 
is not prescribed by the model, it has to be taken from the {\it correlation} 
measurements. Note that $\lambda_*$ is in general rapidity and 
transverse mass dependent and thus may depend on the kinematic 
region where the data are determined. The effective volume factor
$V_*( {x}_s)$ measures the invariant volume element of the 
particle emission at the time of the last interactions.
{\it This factor can be very strongly $m_t$ dependent, 
and its possible $m_t$ dependence should be taken into account 
in the future data analysis}~\cite{tamas2}. This invariant,
 effective volume factor is related to the HBT radius parameters \cite{tamas2}
as
\ben
V_* (x_s) & = 
	& ( 2 \pi \tau_0^2  {\Delta\eta}_*^2)^{{1\over 2}}\,\,
	(  2 \pi R_*^2)  \,\, 
	{\dst \Delta \tau_* \ov \Delta \tau }.
\enn
The effective source dimensions
 in the transverse, longitudinal and temporal directions were
 obtained in Boltzmann approximation in Ref. \cite{tamas2} as 
\ben
{\dst 1 \ov R_*^2 } \, & = & \,
	{\dst 1 \ov R_G^2} \, + \,\,
	{\dst 1 \ov R_T^2 }\, \cosh[ {\eta}_s],  \\
{\dst 1 \ov \Delta \eta_*^2} & = &  
	{\dst 1 \ov \Delta \eta^2 } +
        {\dst 1 \ov \Delta \eta_T^2} \cosh[{\eta_s}] 
	- {\dst 1 \ov \cosh[\eta_s]^2}, \\
{\dst 1 \ov \Delta \tau_*^2 } & = & 
	{\dst 1 \ov \Delta\tau^2} + {\dst 1 \ov \Delta\tau_T^2}
			\cosh[ {\eta}_s] \ , \\
 R_T^2 & = & { \displaystyle\strut \tau_0^2 \over
        \displaystyle\strut a^2 + b^2 }
         { \displaystyle\strut T_0 \over \displaystyle\strut m_t}, 
		\label{e:rt}\\
 \Delta\eta_T^2 & = & {\dst T_0 \ov m_t}, \qquad 
	\hbox{\rm and} \qquad
 \Delta\tau_T^2 \,\, = \,\, {\dst \tau_0^2 \ov d^2} {\dst T_0 \ov m_t}.
		\label{e:tt}
\enn
while the Cooper--Frye flux factor contains 
$n^{\mu}$, the  
normal-pointing unit vector 
of the hypersurface family $d^4\Sigma^{\mu}(x)$.
This Cooper--Frye flux factor is given by
\ben
p^{\mu} n_{\mu}(x_s)  & = &
 	m_t \cosh[ {\eta}_s] 
\enn
In the present status report we do not yet consider the effects
of Bose-Einstein statistics on the saddle-point and on the 
radius parameters of the Bose-Einstein correlation functions.
In {\it Boltzmann approximation}, the saddle point $x_s$ is 
characterized by $\tau_s = \tau_0$,
 $ {\eta}_s =   (y_0 - y)/ [1 +\Delta\eta^2 (m_t/ T_0 - 1)] $,
 $r_{x,s} = \beta_t b R_*^2( {\eta}_s) m_t / (\tau_0 T_0) $
 and $r_{y,s} = 0$, if the saddle-point equations can be
{\it linearized} in the LCMS~\cite{lcms} around $r_x = 0$. 
For the NA44 experiment, the mid-rapidity is
 located at $y_0 = 3$. 

 The same length scales, which determine the effective, invariant volume
factor, $V_*$, shall determine the radius parameters of the BECF.
These length scales, indexed by subscript $*$, are the homogeneity lengths 
in the considered spatial or temporal directions. From the above expressions
it is clear that they are determined by an interplay of the geometrical scales,
$R_G$, $\Delta\eta$, and $\Delta\tau$ and the so-called thermal scales,
$R_T$, $\Delta\eta_T$, and $\Delta\tau_T$, which are indexed with subscript
$T$ and stem from the Boltzmann factor $\exp[ p^{\mu} u_{\mu}(x_s)/T(x_s) ]$.
In Boltzmann approximation, utilizing a solution of the linearized saddle point
equations~\cite{tamas2} one obtains in the mid-rapidity region, for $y = y_0 $,
in the LCMS system~\cite{lcms}:
\ben
	R_{\s}^2 & = & R_*^2, \label{e:side} \\
	R_{\o}^2 & = & R_*^2 + \beta_t^2 \Dt_*^2 ,\label{e:out} \\
	R_L^2 & = & \tau_0^2 \Delta\eta_*^2 \ . \label{e:long}
\enn
The BECF can be expressed as
\begin{equation}
	C(Q;K) = 1 + \lambda_* \exp( - R_\s^2 Q_\s^2 -
			R_\o^2 Q_\o^2 
- R_L^2 Q_L^2 - 2 R^2_{\o,L} Q_\o Q_L ) 
\end{equation}
For the NA44 data, which are sampled in the central region, 
$y\approx 3$, we have $R^2_{\o,L} \approx  0$.
A more detailed discussion on the HBT radii was given in Ref. \cite{tamas2}.

From the above expressions one can figure out that the radius
parameters in general may have complicated dependence on the
rapidity and on the transverse mass of the particle pairs.
This is partly due to the rapidity and transverse mass dependence of the 
$\eta_s = \eta_s^{LCMS}$ variable, the cross-term generating 
hyperbolic mixing angle~\cite{tamas2} defined in the LCMS frame,
and partly due to the simultaneous presence of $m_t$ dependent and
$m_t$ independent terms in eqs.~(\ref{e:rt} - \ref{e:tt}).
We see also, that all radii parameters may be proportional 
to $1/\sqrt{m_t}$ in the case when the total size and 
time spread of the particle emission are much larger than the 
corresponding thermal sizes, $R_{T},\Delta \eta_{T}$
and $\Delta \tau_{T}$, respectively. 

There is no a priori reason to assume that 
$S_{\pi}(x,p) = S_K(x,p)$ i.e. that the emission functions
have the same form  for pions ($\pi$) and kaons ($K$).
Freeze-out arguments are based on the estimated value of the hadronic
cross-sections and the pions have larger cross-sections than the kaons.
On this basis, one would expect that pions will freeze-out later, from a
larger volume, than the kaons.

Our results on the parameters of the two-particle correlation function 
indicate, that even if the geometrical radii for the pion emission function
and the kaon emission function were different, they may play only a
minor role if they are sufficiently large as compared to the thermal
length-scales. Thus the parameters of the Bose--Einstein correlation function
may obey an $m_t$ scaling even if the emission funtions of the various
hadronic pieces were different \cite{footnote}.

The radii parameters are mainly sensitive to the smallest of the 2 scales present,
in contrast to the IMDs which are sensitive to the biggest of the scales.
The model is discussed in detail in Ref. \cite{tamas2}, where the spectrum 
has been calculated in a Boltzmann approximation, allowing for a detailed
analysis of the influence of the thermal and geometrical length scales
on the rapidity width and the slope parameter of the momentum distribution.

\section{Fitting Data with Analytical Formulas and Using Numerical Integrations}

In the previous section, we have discussed some new features 
of the Bose-Einstein statistics on the momentum distribution
obtained in the saddle-point approximation.
In this section, we present a status report about fitting the
{\it Boltzmann} approximations to eq.~(\ref{e:s}) to 
NA44 data, both analytically and numerically.
Our fitted formulas thus are not taken from the previous chapter,
but from Ref.~\cite{tamas2}, which corresponds to an
improved saddle-point approximation to the Boltzmann-approximated
emission function. Work is in progress to study the effects of 
Bose-Einstein statistics on the analytic and numerical fitting of
these NA44 data. In a qualitatively similar analysis of the 
NA49 data the effects arizing from the Bose-Einstein statistics were
reported to be significantly influencing the quality of 
the fit \cite{alcor}.

Thus we have tested the model of ref.~\cite{tamas2}
 by a $\chi^{2}$-fit to the NA44 data on the parameters of the 
kaon and pion BECF as well as on the kaon and pion IMD.
In the fit, the data points for the kaon and pion spcetrum did not include yet
the systematic errors, which were first estimated in Refs.~\cite{murray,xu}.
We have fitted the analytical results for the spectrum obtained in 
a Boltzmann approximation using the modified saddle point method, as described
in Ref.~\cite{tamas2}. 

Our results are still preliminary because of the following reasons:

\begin{description}
\item[{\it i)}] At the time of the fitting the  systematic errors on the
pion and kaon IMD were not yet available, which resulted in 
underestimated  errors on the spectrum.
The small errors in turn imposed a too strong requirement on the fitting,
and increased the relative weight of the spectrum data points with respect
to the radius parameters, for which both the  
systematic  and the statistical errors 
were determined experimentally.  As a consequence, 
the preliminary results discussed below underestimate the
errors of the fitted parameters. 
\item[{\it ii)}] We are currently developing a method which supposedly
will allow us to analytically estimate the errors made when the 
saddle-point integrations are carried out. This could be potentially
very helpful to restrict the domain of the parameters in the analytical 
fitting procedure to that part of the parameter space where the 
analytical results correspond to the numerically integrated 
distributions \cite{num}.
\item[{\it iii)}] We have note yet included the effects of the Bose-Einstein
statistics into the fitting, and finally
\item[{\it iv)}] We currently realized that the $\chi^2$ hypersurface
	has a rather complicated shape when comparing our models to NA44
	pion and kaon data. We hope that the distinction among the
	various almost degenerate but physically different minima can 
	be improved in the near future by including data ondata on  heavier particles
	into the data analysis.
\end{description}
The final results of the data analysis, 
to be obtained by eliminating the problems in items {\it i) - iv)},
will be reported elsewhere.

With the reservations as listed above, we have found the parameter values,
summarized in Table~1, by applying the CERN MINUIT multidimensional fitting
algorithm to determine the lowest values of $\chi^2/NDF$ in the parameter
space of the model. 
When we fit to pion data separately or only to kaon data, separately,
we do not find a fit which were statistically better than
the simultaneous fit to both the pion and kaon sample, see Table 2.  
The overall $\chi^2/NDF$ does not change. 

\begin{center}
\begin{minipage}[t]{11.054cm} 
\noindent
{\small {\bf Table 1.}  
Fitting HBT radius parameters and spectra
 from the model  with analytical approximations, where the domain of the
 validity of the approximations is not taken into account, and with 
 the numerically integrated  radius parameters and spectra.}
\end{minipage}
\vspace{-0.3cm}
\end{center}

\begin{table}[h]
\hskip 1.1cm
\begin{tabular}{lll}
\hline\hline
 \null & Analytical & Numerical \\
 \null & (preliminary errors)  & (preliminary errors)  \\
\hline
$\chi^{2}/ndf$, full fit  & 207/126 & 237/126 \\ 
$\Delta\chi^{2}/bins$, low $p_t$  & 77/42& 104/42\\
$\Delta\chi^{2}/bins$, high $p_t$  & 78/35& 76/35\\
$\Delta\chi^{2}/bins$, kaon & 41/51 & 40/51\\ 
$\Delta\chi^{2}/bins$, radii & 12/9& 18/9\\ \hline
$T_{0}$ [MeV]  & 140 $\pm$ 2& 108 $\pm$ 10\\
$\tau_{0}$ [fm/c]  & 4.6 $\pm$ 0.2&  7.0 $\pm$ 0.5 \\
$R_{G}$ [fm] & 7.0 $\pm$ 1.1& 4.4 $\pm$ 0.2\\
$\D \eta$  & 2.5  $\pm$ 1.3& 1.5 $\pm$ 0.6 \\
$\Delta \tau$ [fm/c] & 3.2$\pm$ 1& 2.4 $\pm$ 1.3\\
$f$ & 0.89 $\pm$ 0.03& $0.88 \pm 0.12$ \\
$b^2$ & 0.52 $\pm$ 0.08& $0.68 \pm 0.14$ \\ 
$d^2$  & 499 $\pm$ 50&  130 $\pm$ 8\\
\hline\hline 
\end{tabular}
\end{table}

 In Fig. 1 we show the result of the fit with the analytical expressions,
 displayed by a dashed line. The measured data points together with their
error bars are also shown.
Parameters $a$ and $b$ appear in a combined manner
in the analytic expressions for the invariant momentum distribution, that
is why the new variable $  f  =  { b^2 / (a^2 + b^2 )}$ is introduced 
in the fit.

The analytical formulas give a slightly better adjustment to the data than the
description obtained by the numerically integrated emission function.
This may either indicate that the analytical fit develops a fake minimum,
or alternatively that some physics appears effectively in the 
analytic approximations which was still missing from the emission function
of eq.~(\ref{e:s}).
The analytical formulas thus correspond to the source given by
eq.~(\ref{e:s}) under certain restrictions only.
A detailed numerical study indicated \cite{num} that the conditions 
$$\frac{r_{x,s}}{\tau_0}=
\frac {\beta_{t}bR_{*}^{2}}{\tau_{0}^{2}\Delta\eta_{T}^{2}}<0.6$$
 and $\Delta\eta_*(y,m_t) < 0.9$ 
have to be satisfied simultaneously in order to reach less than 
20 \% relative error in the HBT
radius parameters and the slope parameter of the IMD.
This condition is violated by our analytic  fit to the pion sample:
from $\frac{r_{x,s}}{\tau_0} < 0.6 $ we obtain $ m_{t,\pi} < 200 $ MeV,
while from $\Delta\eta_* < 0.9$ one obtains $m_{t,\pi} > 300 $ MeV,
and these conditions cannot be satisfied simultaneously.
The same conditions imply $m_{t,K} < 600 $ MeV,
which is satisfied for the NA44 kaon data.

\begin{center}
\begin{minipage}[t]{11.054cm} 
\noindent 
{\small  {\bf Table 2.} 
 Fitting either NA44 $\pi$ spectrum  and radius parameters
or $K$ radius parameters and spectrum,  with the numerically
integrated model, in Boltzmann approximation.}
\end{minipage}
\vspace{-0.4cm}
\end{center}

\begin{table}[h]
\hskip 1.1cm
\begin{tabular}{lll}
\hline\hline
 & Only $\pi$ & Only $K$ \\
 & (preliminary errors)  & (preliminary errors)  \null \\
\hline
$\chi^{2}/ndf$, full fit  & 195/76 & 40/48   \\ 
$\Delta\chi^{2}/bins$, low $p_t$  & 103/42& --\\
$\Delta\chi^{2}/bins$, high $p_t$  & 75/35& --\\
$\Delta\chi^{2}/bins$, kaon & -- & 40/51\\ 
$\Delta\chi^{2}/bins$, radii & 16/6 & 0/3\\ \hline
$T_{0}$ [MeV] & 112 $\pm$ 2 & 133 $\pm$ 6\\
$\tau_{0}$ [fm/c]  &  6.6 $\pm$ 0.3 & 6.4 $\pm$ 0.4\\
$R_{G}$ [fm] & 4.1 $\pm$ 0.1 & 12 $\pm$ 1\\
$\D \eta^2$  &  =2.2  & =2.2\\
$\Delta \tau^{2}$ [fm$^{2}]$ & =5.5 & =5.5\\
$f$ & $ 1.0 \pm 0.2$ & 0.52 $\pm$ 0.1\\
$b^2$ & $0.532 \pm 0.02$ & 0.58 $\pm$ 0.02\\ 
$d^2$ &  =130 & =130 \\
\hline\hline 
\end{tabular}
\end{table}

Note that the restrictions given by the IMD data points,
which do not yet have the systematic errors, seem to be so strong that
 the HBT $m_t$-dependent radius parameters from the fit 
are forced to deviate from the $R_L \simeq 
R_\s \simeq R_\o \propto 1/\sqrt{m_t}$ rule, which is well satisfied by the 
NA44 data points. If we do tried not to fit the IMD of pions and kaons, our
model were able to perfectly fit the 9 NA44 HBT data points, with large allowed
region for the geometrical source sizes. 

On this basis we expect that the inclusion of the systematic errors for the
IMD of pions and kaons will improve the agreement between the model and the 
data. Note, however, that  a fit to the measured particle spectrum is rarely
obtained in such a good quality as obtained here.
The fit with numerical integration is shown on Fig. 1 with a full line. 
The $\chi^2$ of the fit is somewhat high but still acceptable.
The fit with the analytical formulas is shown on Fig. 1 with a dashed line.
The difference between the two fits is only seen in the low $p_t$ region
for the pions.

\begin{center}
\vspace*{15.5cm}
\includegraphics{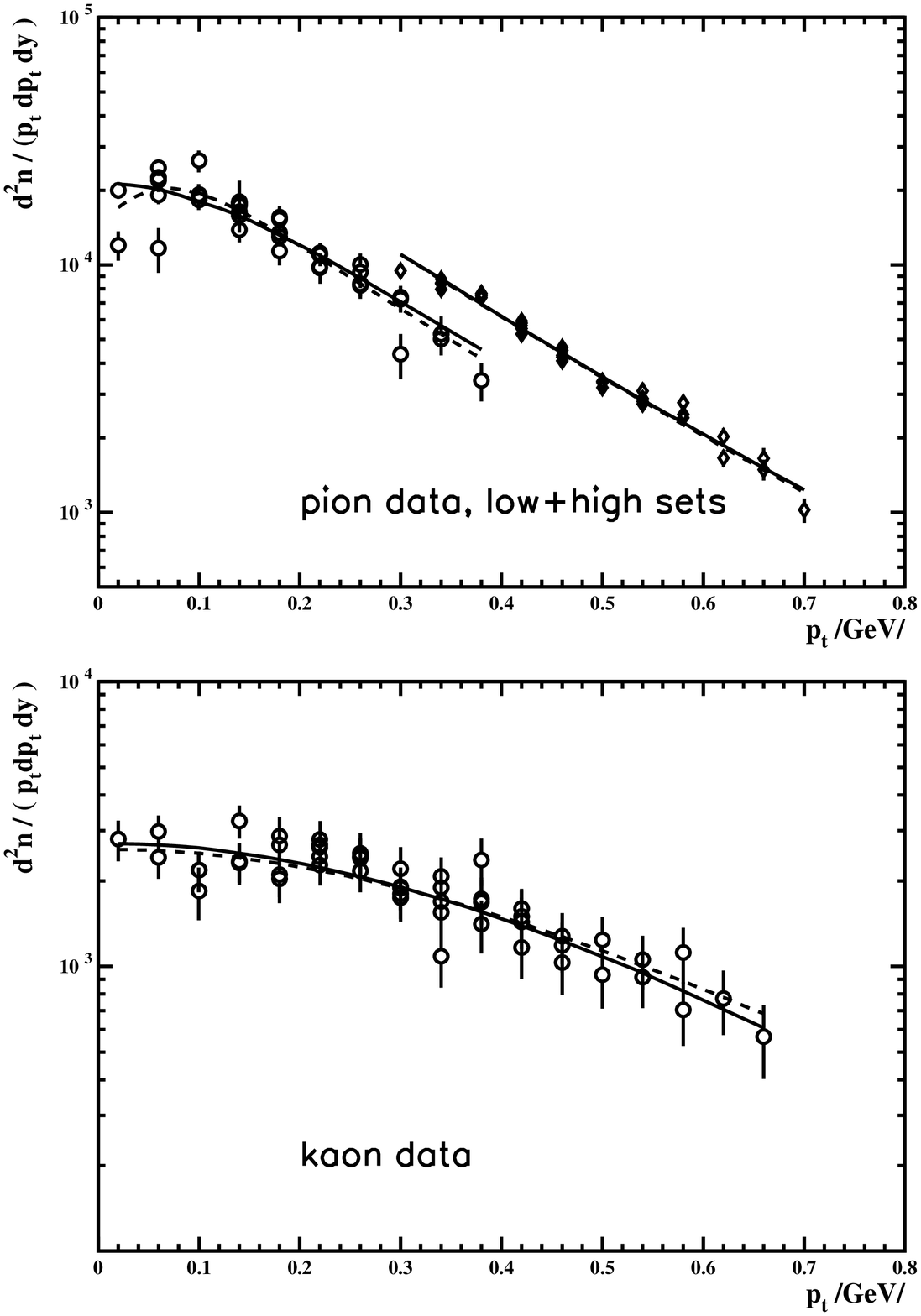}
\vskip -70pt
\begin{minipage}[t]{11.0540cm}
\noindent
 {\small {\bf Fig.~1.} 
The upper half shows the two pion IMD, 
the lower half the kaon IMD. Data are here
displayed versus $p_t$ and they are unnormalized. 
The full line shows the result of the fit with numerical integration, 
the dashed line the result using the analytical
formulas. The rapidity of the data points is not shown. 
The acceptance is restricted to a small region in rapidity 
which changes with $p_t$ specially for the
pion data \cite{na44mt}. More than one data point for a 
$p_t$ means  neighbouring rapidity bins. The rapidity of the pions varies 
from 4.0 to 2.6, the rapidity of kaons
spans 3.4 to 3.1. Data points at the edge of acceptance are not used.}
\end{minipage}
\end{center}
\vskip 4truemm

\begin{center} 
\vspace*{15cm}
\includegraphics{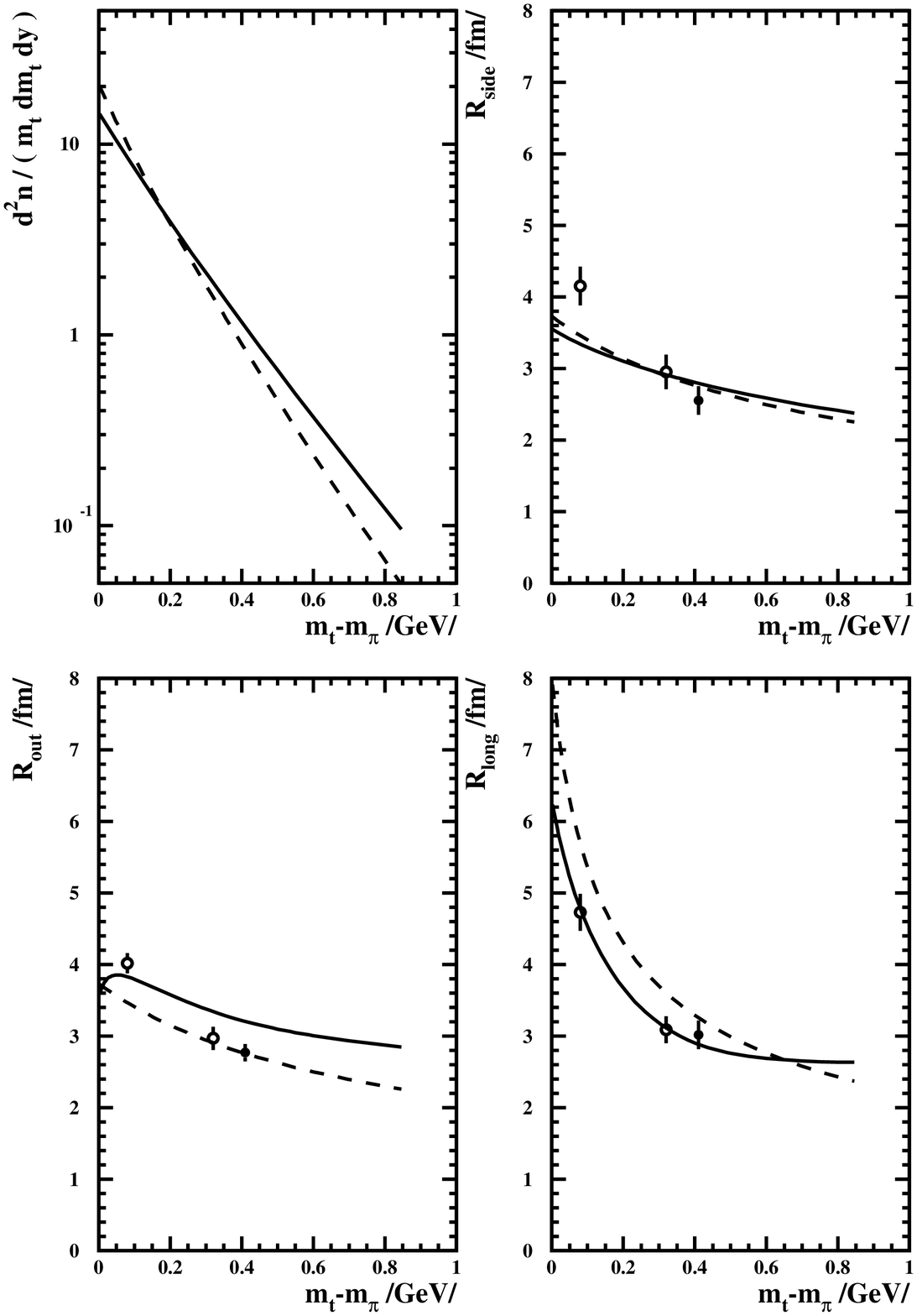}
\vskip -70pt
\begin{minipage}[t]{11.054cm}
{\small {\bf Fig.~2.}  
Full line shows the fit using the numerical integration, dashed line shows
the result of the analytical formulas with the parameters of the same fit.
Data points are the radii measurements of NA44 \cite{na44mt}. Unfilled
circles are used for pion data, filled for kaon data.}
\end{minipage}
\end{center}
\vskip 4truemm

Fig. 2 shows the fit with numerical integration to the radii parameters
and also the difference between the numerical integration and 
the analytical calculation using the parameters of the fit with numerical 
calculation. 
We see important differences mainly for the IMD.
The radius parameters  are reproduced by the analytic expressions with less
than 20 \% relative error at the transverse mass of the data points,
and the slope parameters of the IMD for the pions is also reproduced up to
a 20 \% relative error.

When we fit the analytical formulas without restricting their domain
of validity, the results are given in Table 1.
Unfortunately,
those parameter values  which correspond to the best description of the data
by the analytical formulas  are outside of that region of the parameter space,
 where the analytic approximations correspond to the numerically evaluated
emission function.  Either the model emission function needs to be
further adjusted in such a manner that its numerically integrated spectrum and
HBT radius parameters will closely correspond to the presently
found analytical formulas, or, if the emission function is fixed
at its present form, one should not allow for fitting the data with the 
analytical formulas in an unrestricted domain of the parameter space. A
more detailed analysis is needed, which can take
 into account that the errors of the analytic formulas increase strongly
 if we leave the domain of the 20\% relative errors. 
This domain is given by the requirements
$\frac{r_{x,s}}{\tau_0} \leq 0.6$ and
$\Delta\eta_* < 0.9$, see Ref.~\cite{num} for further details.

The fit implies that the model parameters $a$, $b$ and $d$ 
take  values significantly
different from 0,
approximately 0.3, 0.8 and 11, respectively.

\section{Conclusions}
	This hydrodynamical model gives a fair description of data.

	The analytical approximations provide less than 20\% relative error
	for the HBT radius parameters and the slope parameters of the spectra
	in that part of the parameter space where the NA44 data are.

	If we fit the approximative analytic results to data without estimating
	the errors of the approximations in the region,
	 where the fit is performed, we find a fake minimum, where the 
	extrapolated analytic results give a very good description of the data.
	However, at this part of the parameter space, no minimum is found when
	numerical integration is invoked. 
	We find no evidence for different model parameters 
	for pions or for kaons.

	In a preliminary analysis, we
	find a surprisingly low central temperature at the 
	mean time of the last interactions, \mbox{$T_0 = 108 \pm 10 $ MeV}.
	This is in agreement with our preliminary results presented at
	Quark Matter'95~\cite{qm95}. Note that a similar low value of
	\mbox{$T_0 = 92.9 \pm 4.4$ MeV} has 
	been found at AGS energies from a detailed analysis of the 
	BECF and the spectrum of both pions and kaons~\cite{chapman}. 
	However, we would like to be careful and not to draw very strong 
	conclusions untill we can find a very well defined minimum
	by adding more data, utilizing the experimental estimate of the 
	systematic errors on the spectrum, and untill we have included
	the effects of quantum statistics  on the single-particle
	spectra into the data analysis.
 
\vskip 10pt
\section*{Acknowledgement}
One of us (B.L) wants to thank the organizers for the invitation to this very
stimulating workshop.
 
\section*{Notes} 
\begin{itemize}
\item[\null]
a. E-mail: csorgo@sunserv.kfki.hu
\item[\null]
b. E-mail:  bengt@quark.lu.se
\end{itemize}


\vfill\eject 

\begin{thebibliography}{99}\parindent=8truemm
\itemsep -1mm
\bibitem{tamas2}
      T. Cs{\"o}rg\H o and B. L{\"o}rstad: hep-ph/9509213,
	 {\it Phys. Rev. C} {\bf 54} (1996) 1390.

\bibitem{na44mt} 
      H. Beker et al., NA44 Collaboration,  
	{\it Phys. Rev. Lett.} {\bf 74} (1995) 3340.

\bibitem{michael}
	Michael Murray, NA44 Collaboration, private communication.

\bibitem{halo} T. Cs\"org\H o, B. L\"orstad, J. Zim\'anyi,
	hep-ph/9411307, {\it  Z. Phys. C} {\bf 71} (1996) 491.

\bibitem{schlei}  J. Bolz, U. Ornik, M. Pl\"umer, B. R. Schlei 
	and R. M. Weiner,  \hfill\break
	{\it Phys. Rev.} {\bf  D47} (1993) 3860.

\bibitem{uli}	S. Chapman, P. Scotto and U. Heinz,
	{\it Heavy Ion Physics} {\bf 1} (1995) 1;\hfill\break
	 {\it Phys. Rev. Lett.}  {\bf 74} (1995) 4400.

\bibitem{qm95}  T. Cs\"org\H o and B. L\"orstad,
	{ \it Nucl. Phys.} {\bf A590} (1995) 465c.

\bibitem{akk}  S. V. Akkelin and Yu. M. Sinyukov,
	Bogoljubov Institute Report No. ITP - 63 -94E, December 1994
	(unpublished).

\bibitem{lcms}	T. Cs\"org\H o and S. Pratt, 
	Report No. KFKI - 1991 -28/A, p. 75 (1991).

\bibitem{footnote} {This was not seen in Ref.~\cite{us96},
where it was assumed that pions and kaons are characterized by the same
emission function}.

\bibitem{murray} M. Murray, NA44 Collaboration,
	{\it	 Heavy Ion Physics} {\bf 4} (1996) 213. 

\bibitem{xu}	N. Xu, NA44 Collaboration,
	{\it Heavy Ion Physics} {\bf 4} (1996) 263.

\bibitem{num}   T. Cs\"org\H o, P. L\'evai, B. L\"orstad,
	hep-ph/9603373, \hfill\break
	{\it Acta Phys. Slovaca} {\bf 46} (1996) 585.

\bibitem{chapman} S. Chapman and J. Rayford-Nix, Report No. LA-UR-96-0782.

\bibitem{us96}	U. Heinz, B. Tom\'asik, U. A. Wiedemann and Wu Y.-F.,
	\hfill\break
	{\it Heavy Ion Physics} {\bf 4} (1996) 249.

\bibitem{alcor} J. Zim\'anyi, T. S.  Bir\'o, T. Cs\"org\H o, P. L\'evai,
	{\it		Heavy Ion Physics} {\bf 4} (1996) 15.



\end{thebibliography}
\end{document}